\documentclass[ aip,
amsmath,amssymb,
reprint,%
]{revtex4-1}

\usepackage{graphicx}

\usepackage{dcolumn}
\usepackage{bm}

\usepackage[utf8]{inputenc}
\usepackage[T1]{fontenc}
\usepackage{mathptmx}

\begin{document}

\preprint{AIP/123-QED}

\title{Skyrmions in antiferromagnets: thermal stability and the effect of external field and impurities} 

\author{Maria N. Potkina}
\affiliation{Science Institute and Faculty of Physical Sciences, University of Iceland, 107 Reykjav\'{\i}k, Iceland}
\affiliation{Dept. of Physics, St. Petersburg State University, St. Petersburg, 198504, Russia}
\affiliation{ITMO University, 197101 Saint Petersburg, Russia}

\author{Igor S. Lobanov}
\affiliation{Dept. of Physics, St. Petersburg State University, St. Petersburg, 198504, Russia}
\affiliation{ITMO University, 197101 Saint Petersburg, Russia}

\author{Hannes J\'onsson}
\email{hj@hi.is}
\affiliation{Science Institute and Faculty of Physical Sciences, University of Iceland, 107 Reykjav\'{\i}k, Iceland}
\affiliation{Dpt. of Applied Physics, Aalto University, FI-00076 Espoo, Finland}

\author{Valery M. Uzdin}
\affiliation{Dept. of Physics, St. Petersburg State University, St. Petersburg, 198504, Russia}
\affiliation{ITMO University, 197101 Saint Petersburg, Russia}



\begin{abstract}
Calculations of skyrmions in antiferromagnets (AFMs) are presented, and their properties compared with skyrmions in corresponding ferromagnets (FMs). The rates of skyrmion collapse and escape through the boundary of a track, as well as the binding to and collapse at a non-magnetic impurity, are calculated as a function of applied magnetic field. The activation energy for skyrmion annihilation is the same in AFMs and corresponding FMs in the absence of an applied magnetic field. The pre-exponential factor in the Arrhenius rate law is, however, different because skyrmion dynamics is different in the two systems. An applied magnetic field has opposite effects on skyrmions in the two types of materials. In AFMs the rate of collapse of skyrmions as well as the rate of escape through the edge of a magnetic strip decreases slightly with increasing field, while these rates increase strongly for a skyrmion in the corresponding FMs when the field is directed antiparallel to the magnetization in the center of the skyrmion. A non-magnetic impurity is less likely to trap a skyrmion in AFMs especially in the presence of a magnetic field. This, together with the established fact that a spin polarized current moves skyrmions in AFMs in the direction of the current, while in FMs  skyrmions move at an angle to the current, demonstrates that skyrmions in AFMs have several advantageous properties over skyrmions in FMs for memory and spintronic devices.
\end{abstract}


\maketitle


\section{Introduction}

Topological magnetic chiral structures are of interest as a manifestation of topology in fundamental physics, as well as in connection with applications
such as the development of a new generation of magnetic storage and data processing 
devices.\cite{Fert17,Buttner18,Everschor-Sitte18}
Magnetic skyrmions have been observed in various thin layers of ferromagnets (FMs)  adsorbed on a surface of a heavy metal,\cite{Heinze2011} 
in multilayer systems \cite{Woo2016} and in materials with broken inversion symmetry.\cite{Tokunaga15}
Skyrmions can be characterized by 
topological charge, which is an integer that cannot be changed by a continuous transformation of magnetization. 
A skyrmion state and a ferromagnetic state have different topological charges and therefore cannot be transformed from one to another in a continuous medium.
In that sense, a skyrmion is topologically protected.
However, when the magnetic moments are localized at sites of a discrete lattice, there is a finite activation barrier for the collapse of a skyrmion 
and its lifetime becomes shorter the higher the temperature is.\cite{Uzdin18-JMMM,Bessarab2018}

Skyrmions could be used as magnetic bits in racetrack memory with high density, speed and energy efficiency. 
However, small skyrmions 
in FMs
with diameter of 10 to 20 nm have been found to be stable only at low temperature \cite{Wiesendanger16} while skyrmions that are stable at room temperature are large, 
with diameter exceeding 50 nm.\cite{soumyanarayanan2017} 
The stray field of large skyrmions leads to complicated skyrmion interactions and sets a limit to their density. 
Skyrmions can be moved by a spin polarized current but in 
a FM the direction of movement is at an angle to the current, the so-called Hall angle.\cite{Litzius2017} 
This can lead to the escape of skyrmions through the boundary of the track. 
Some of these problems may be overcome by using other materials or topological magnetic structures. 
Accordingly, skyrmions in antiferromagnets (AFMs) and ferrimagnets,\cite{Smejkal18,Gomonay2016} as well as antiskyrmions,\cite{Nayak2017}
have recently attracted increased attention. 

Skyrmions in AFMs can have several advantages over skyrmions in FMs
but they are hard to detect.
Recently, however,
skyrmions that are stable at room temperature have been observed in a
synthetic antiferromagnet based on Pt/Co/Ru multilayers.
This system consist of several FM layers with antiparallel magnetization separated by metal spacers. 
The skyrmions in this type of structure have been detected using magnetic force microscopy.\cite{Legrand19}
Skyrmions have also recently been observed in ferrimagnetic materials.\cite{Caretta18,Woo18}
At a certain temperature, the magnetization of the two sublattices 
of the ferrimagnet
is compensated and the properties of the skyrmions are then
similar to the properties of a skyrmion in an AFM. 
Below and above the compensation temperature, skyrmions in a ferrimagnet can be detected by the same methods as in FM. 

The observation and control of skyrmions in AFM materials is a challenging task that has to be solved for their utilization. 
The magnetic moment associated with a skyrmion in an AFM is small and methods typically used to register skyrmions in FM materials do not work. 
The core of a skyrmion in an AFM looks like the AFM material outside it but the domain wall at the boundary can be detected for example using spin polarized scanning tunneling microscopy because the direction of the magnetic moments inside the domain wall is different. 
AFM skyrmions can be obtained in ferrimagnetic materials near the temperature of compensation.\cite{Caretta18} 
Below and above this temperature, the skyrmion posses some magnetic moment which can be measured.
Therefore, local perturbation such as a variation in the temperature can make the skyrmion visible. 
Also, sublattices in ferrimagnets contain atoms of different chemical elements. 
Therefore, element specific methods such as element-specific scanning transmission X-ray microscopy\cite{Woo18} 
can be used to detect topological structures without net magnetization.
Finally, topological Hall effect for AFM and FM skyrmions is quite different and could be used for detection and manipulation of skyrmions.
\cite{Buhl17}

Theoretical arguments for the existence of vortex structures in easy-axis gyrotropic antiferrimagnets were given by Bogdanov {\it et al.}  
within the framework of continuum micromagnetic theory.\cite{Bogdanov89,Bogdanov99}
Skyrmion-like objects in continuum models were also predicted in a certain subclass of antiferromagnetic materials.\cite{Bogdanov02}

More recently, theoretical analysis of skyrmions in a continuum representation of an AFM material has been presented by Keesman {\it et al}.\cite{Keesman2016}
The ground state was determined  
using Monte Carlo simulations and the results compared to theoretical analysis. 
For finite systems, a phase that shows similarities with skyrmions was identified in between the spin flop and the spiral phases.

Zhang {\it et al.}\cite{Zhang16} carried out micromagnetic lattice simulations and mapped out a 
phase diagram for AFM skyrmions as function of the exchange stiffness, Dzyaloshinskii-Moriya (DM) and perpendicular anisotropy constants,
and demonstrated two approaches for creating an AFM skyrmion. 
Jin {\it et al.}\cite{Jin16} analyzed the dynamics of a skyrmion in AFM and showed that the minimum driving current density is about two orders of magnitude 
smaller than for a FM skyrmion, the velocity is significantly larger and the movement is in the direction of the current 
(while skyrmions in FM move at an angle with respect to the current). 
Theoretical studies have also been carried out of 
the dynamics of AFM skyrmions in the presence of an inhomogeneous distribution of magnetic anisotropy,\cite{Shen18}
temperature gradients,\cite{Liang19}
and defects.\cite{Khoshlahni19}

Barker and Tretiakov, and Bessarab {\it et al.}\cite{Barker16,Bessarab19} analyzed the dynamics of a skyrmion in AFM 
as well as the effect of temperature on its size and phase stability.
The activation energy for collapse of the skyrmion was evaluated and 
the lifetime of the skyrmion estimated
to be on the order of milliseconds for a temperature range of 50 K to 65 K and a certain set of parameter values.
However, from these calculations it could not be concluded whether compact AFM skyrmions are possible at room temperature.


In the present article, the properties of skyrmions in AFMs are compared with the properties of skyrmions in corresponding FMs. 
By reversing the sign of the exchange and DM parameters and flipping every other spin, one system is mapped onto the other and
the energy surfaces characterizing these two systems are shown to be identical in the absence of a magnetic field.
The extensive knowledge that has already been obtained on skyrmions in FM can, therefore, be used to predict properties of skyrmions in AFMs. 
The activation energy for the various skyrmion annihilation processes is the same in the absence of an applied field, 
but the dynamics of skyrmions in AFMs is different from that in FMs
so the pre-exponential factor in the Arrhenius rate law is different.
In the presence of an applied magnetic field, the activation energy for the annihilation of a skyrmion in AFM is close to that of a skyrmion 
in the corresponding FM with modified anisotropy parameter and zero field. 
A non-magnetic impurity tends to bind the skyrmion and increase the rate of collapse, but the effect of an applied magnetic field is opposite for AFM and
the corresponding FM. Numerical calculations of these processes are presented below for a characteristic AFM film. 


\section{Model}

We consider a system of magnetic moments localized at the nodes of a square lattice track in the x-y plane.
The width of the track is 60 lattice sites along the y-direction with free boundary conditions at the edges.
The simulation cell contains also 60 lattice sites along the x-direction but with periodic boundary conditions applied there.
In this two-dimensional (2D) model, each spin represents a column of spins, with effective parameters chosen to represent a three-dimensional film.
 
The energy of the system is given by a generalized Heisenberg model:
$$ 
E \ = \ -  \sum_{<i,j>} (J_{ij}{\bf S}_i \cdot {\bf S}_j +{\bf D}_{ij}\cdot ({\bf S}_i \times {\bf S}_j)) $$
\begin{equation}
  \ \ \ \ \ \  - \sum_j ( K ({\bf S}_j \cdot {\bf e}_z)^2 + \mu {\bf B} \cdot {\bf S}_j )
\end{equation}
where $ {\bf S}_j $ is a unit vector pointing in the direction of the magnetic moment on atom $ j $, 
the sum over $<i,j>$ is taken over all pairs of nearest neighbor atoms, 
$J_{ij}$ is the Heisenberg exchange parameter, 
${\bf D}_{ij}$ is the DM vector, 
$K$ is the constant of easy axis anisotropy (along z-direction, $K>0$), 
$\bf B$ is applied magnetic field 
with magnitude $B=|{\bf B}|$
and $\mu$ is the magnitude of the magnetic moments.
For FM $ J_ {ij}> 0 $ and for AFM $ J_ {ij} <0 $. 

The Hamiltonian used here is equivalent to the one used by Zhang {\it et al.}
\cite{Zhang16} 
in their study of the phase diagram and 
in the calculations presented below we choose parameter
values that are representative for a metastable skyrmion state:
$|J| =4.97 \cdot 10^{-21}$~J, $K=0.8 \cdot 10^{-21}$~J, $D=1.76 \cdot 10^{-21}$~J
(or $|J|$ = 31 meV, $K$ = 5 meV, $D$ =  11 meV).
This corresponds to $D/|J| \approx $ 0.35, $K/|J| \approx $ 0.16.

Note that dipole-dipole interaction is not included in the Hamiltonian. For skyrmions in FM with a radius 50 nm or larger this interaction can 
play a significant role\cite{Buttner18} but can,
to first approximation, be taken into account effectively by modifying the anisotropy.\cite{Lobanov2016} 
However, for AFM the dipole-dipole interaction is almost completely suppressed and can be neglected even for micron-scale structures.

The AFM lattice can be divided into two square checkerboard-like sub-lattices labeled $a$ and $b$ 
and can be mapped into the FM lattice by reversing the direction of magnetic moments in one of the sub-lattices
${\bf S}_i^a \mapsto - {\bf S}_i^a$.
If at the same time the Hamiltonian parameters are changed as 
$J_{ij} \mapsto - J_{ij}$, ${\bf D}_{ij} \mapsto - {\bf D}_{ij}$
the AFM is mapped into a corresponding FM with the same energy if there is no applied  
magnetic field, $ B=0$. 
The energy surface, i.e. the variation of the energy as a function of the orientation of all the spins, 
is exactly the same for the AFM as for the corresponding FM.

The effect of an external magnetic field on the transformation between a skyrmion in FM and a skyrmion in AFM 
has been presented on the basis of a continuum model.\cite{Keesman2016} 
Denoting by ${\bf L}$ the staggered magnetization and by ${\bf M}$ the total magnetization,
the energy density $\omega ({\bf r})$ for AFM can be written as:
$$\omega({\bf r})=-\frac{\mathcal{J}}{2} ( \|{\bf \nabla L}\|^2+8 \|{ \bf M}\|^2)-
M_s
BM_z -\mathcal{K}{L_z}^2  + \mathcal{D} {\bf L} \cdot \mathrm {rot}{\bf L}   $$
Here, $\mathcal{J}, \ \mathcal{K}$ and $\mathcal{D}$ are the densities corresponding to the energy parameters in Eqn. (1), and $M_s$ is the magnetization.
The continuous vector fields ${\bf L}$ and ${\bf M}$ can be
related to the spin orientation ${\bf S}_j$ on the lattice
as follows: ${\bf L}=({\bf S}^a-{\bf S}^b)/2$, ${\bf M}=({\bf S}^a+{\bf S}^b)/2$. 
Here, ${\bf S}^a$ and ${\bf S}^b$ correspond to the values of $\bf S$ on the two sub-lattices.

By minimizing $\omega ({\bf r})$ with respect to $\bf M$ under the condition ${\bf M}\cdot{\bf L}=0$, 
the following expression for the energy density of a (meta)stable skyrmion state is obtained as a function of only the staggered component:
$$ \omega({\bf r})= -{\mathcal{J} \over 2}  \|{\bf \nabla L}\|^2 + \frac{(M_s B)^2}{16 |\mathcal{J} |} (L_z^2-1) -\mathcal{K}{L_z}^2  
+\mathcal{D}{\bf L} \cdot \mathrm {rot}{\bf L}    $$
When $B=0$, this expression coincides with $\omega ({\bf r})$ for FM with magnetization $\bf L$. 
If ${\bf B}\ne 0$ the staggered magnetization in an AFM state coincides with the magnetization in a FM state without magnetic field but with modified anisotropy 
\begin{equation}
 K \mapsto K-{{(M_s B)^2} \over {16 | \mathcal{J} |}}
 \end{equation}
Therefore, there is a one-to-one correspondence between (meta)stable states in AFM and in the corresponding FM.
The energy of a skyrmion in AFM is affected by the application of an external magnetic field in the same way as the change in the anisotropy constant given by Eqn.~(2).
%
While the transformation described by Eqn.\,(2) is obtained for a continuum model, it also gives a good approximation for a discrete, atomic scale model, 
not only for the skyrmion state but for the whole path for skyrmion annihilation, as will be demonstrated below.

Both directions of the field perpendicular to the plane of the lattice are equivalent for a skyrmion in an AFM, 
but they lead to a different effect on a skyrmion in an FM. 
If the field is directed along the magnetization of the ferromagnet, it reduces the radius of the FM skyrmion.
In the calculations presented here, this will be the direction of the applied magnetic field.
If the field is directed in the opposite way, the size of the FM skyrmion increases and it easily becomes unstable with respect to a spiral structure.
In the system studied numerically below, a field of just $\mu B/|J|=-0.0012$ is sufficient to produce spiral structure.
In experimental studies of skyrmions in FMs, the field is invariably directed along the magnetization of the ferromagnet.



\section{Methodology}

The focus of the work presented here is an analysis of the various annihilation processes of skyrmions in AFMs and comparison with skyrmions in
corresponding FMs.  Also, the binding to and dissociation from a non-magnetic impurity is calculated. 
The energy as a function of all the variables representing degrees of freedom in the system, for example the angles describing the orientation of the magnetic moments, 
is referred to as the energy surface characterizing the system. 
We, however, use Cartesian coordinates of the magnetic moments and Lagrange multipliers to keep the magnitude 
of each magnetic moment 
constant in the calculations presented below.
Metastable states are characterized by local energy minima on this energy surface and the stable state by the global minimum.
A transition from one state to another is characterized by the minimum energy path (MEP) connecting the initial and final state minima.
The MEP for a magnetic transition can be found using the geodesic nudged elastic band method.\cite{Bessarab2015}
For large systems with many degrees of freedom it can be helpful to
take advantage of previous knowledge of the saddle point and the shape of the MEP in its vicinity to reduce the computational effort.\cite{Lobanov2017}
The activation energy for a transition can be estimated from the highest energy along the MEP. 
A maximum along the MEP corresponds to a first order saddle point on the energy surface.\cite{Bessarab13} 
 
The rate of a transitions in a magnetic system, such as a skyrmion annihilation process, can be estimated using 
the harmonic approximation to transition state theory (HTST) for magnetic systems.\cite{Bessarab2012,Bessarab13} 
The transition state is then taken to be a hyperplane going through the first order saddle point 
with normal vector pointing along the direction of the eigenvector corresponding to the negative eigenvalue of the Hessian at the saddle point, $H^{sp}$.
The energy surfaces defined by the Hamiltonian in Eqn.~(1) for an AFM material and the corresponding FM material 
coincide in the absence of an applied magnetic field. The activation energy for collapse is, therefore, the same for the two types of skyrmions if $B$=0.


HTST predicts an Arrhenius-type rate law for transitions between magnetic states
%
$$k = \nu_0 \exp\left(\frac{-\Delta E}{k_BT}\right) .$$ 
The activation energy, $\Delta E$, equals the energy at the (highest) maximum along the MEP minus the energy at the minimum corresponding to the initial state. 
The pre-exponential factor 
can be written as\cite{Bessarab2012}
\begin{equation}
\nu_0= \ \frac{\lambda \Omega_0}{2\pi}
\end{equation}
where the factor $ \Omega_0 $ is a ratio of determinants of the Hessian at the initial state minimum, $ H^{min} $, and at the saddle point, $ H^{sp} $,
i.e. 
\begin{equation}
\Omega_0 = {{ \sqrt{\det{H^{min}} } }\over {\sqrt{ \left| \det{H^{sp}} \right|}}}.
\end{equation}
It is connected with the ratio of the entropy of the initial state and the transition state. 
The factor $ \lambda $ is connected with dynamics of the system and relates to the flux through the transition state.
An explicit expression for the rate constant has previously been given in
terms of eigenvalues and eigenvectors of the Hessian at the saddle point.\cite{Bessarab13,Bessarab2012}
%
%
%
But, here $\lambda$ is written in a basis invariant form as
\begin{equation}
\lambda=\sqrt{\frac{{\bf b}\cdot H^{sp} {\bf b}}{|\zeta|}},\quad {\rm with} \ \ 
{\bf b}=\frac{\gamma\zeta}{\mu} {\bf S}^{sp} \times {\bf e},
\end{equation}
where ${\bf S}^{sp}$ is the spin configuration at the saddle point
and 
$\bf e$ is the eigenvector corresponding to the negative eigenvalue of $H^{sp}$ 
(a unit tangent vector to the MEP at the saddle point).
This expression for $\lambda$ is easier to evaluate numerically 
since it can be computed in spin-related basis without the evaluation of
the eigenvectors of $H^{sp}$.
Also,
$\lambda$ can then be interpreted without
referring to the dividing surface.
Indeed, ${\bf b}\cdot H^{sp} {\bf b}$ is the value of the Hessian in the direction $\bf b$.
The velocity is zero at ${\bf S}^{sp}$ since the gradient is zero, but the velocity at other points along the MEP is not zero,
and in the vicinity of the saddle point the velocity
is equal to $\bf b$ times the distance from the saddle point.
This means that 
the value of ${\bf b} \cdot H^{sp}{\bf b}$ can be interpreted as the curvature of the
energy surface at the saddle point in the direction of $\bf b$.

When several different transitions are possible from a given initial state, as for example the different annihilation processes considered here for the skyrmion,
then the lifetime of the initial state can be obtained from the inverse of the sum of the rate constant of the various transitions
$ \tau= 1/\sum_i k_i$.
The lifetime of the skyrmion can thus be obtained from the sum of the rate of collapse (i.e. annihilation within the track) and escape through the boundary of 
the track.  In the presence of an impurity, there are additional processes such as binding to the impurity and possible collapse there or detachment from
the impurity. The time evolution of such a system where multiple possible transitions take place can be simulated using a kinetic Monte Carlo approach. 

The harmonic approximation in HTST does not apply to degrees of freedom for which the energy is nearly constant. 
For example, a skyrmion can translate almost freely within the track for the system studied here. 
Such degrees of freedom need to be handled separately in the rate expression.
They are often referred to as zero modes.
An integration along these modes needs to be performed to obtain the corresponding entropy.\cite{Ivanov17,Bessarab2018} 
This gives the volume of the subspace corresponding to the zero-modes. 
As a result, $ \Omega_0$ has an additional factor 
$Z= (2 \pi k_B T)^{\frac{n_{min}-n_{sp}}{2}}~V_{sp}/V_{min}$, 
where $n_{sp}$, $V_{sp}$ and $n_{min}$, $V_{min}$ give the number of zero modes and the corresponding
volumes at the saddle point and at the initial state minimum, respectively.
We make here the assumption that the skyrmion can translate freely inside the track and also at the saddle point for collapse
(i.e. there are two zero modes in both the initial state and transition state, n$_{min}$=n$_{sp}$=2 ), 
while it can translate freely only in one direction at the saddle point for escape (n$_{min}$=2, n$_{sp}$=1).

The transformation of exchange $J \to -J$ and DM interaction $D \to -D$ with simultaneous change of anisotropy and external magnetic field in accordance with 
Eqn.~(2) and $B \to 0$ conserves the shape of energy surface. Therefore, the entropy factor $\Omega_0$ is the same for the skyrmion collapse in AFM and
the corresponding FM. However $\lambda$ which is connected with the linearized  Landau-Lifschitz equation in the neighborhood of the saddle point 
is different for skyrmions in AFMs and the corresponding FMs, as illustrated below.

We note that in an alternative approach for estimating the transition rate based on the Kramers-Langer approximation,\cite{Coffey01}
$\lambda$ is the positive eigenvalue of the Landau-Lifschitz-Gilbert equation, linearized around the saddle point. 
It characterizes the unstable barrier-crossing mode.\cite{Desplat18}


\section{Results}

We now present results of numerical calculations for the model and parameter values specified above.

\subsection{Effect of applied field on skyrmion size}

Fig.~1 shows how an applied magnetic field affects the radius of a skyrmion in the AFM and in the corresponding FM. 
The radius is defined here as the distance from the center to the region where the magnetic moments have no out of plane component. 
The size of the skyrmion changes in opposite ways in the two cases. 
While the radius of the skyrmion in the FM decreases when the field is directed against the moment in the center of the skyrmion,
the radius of the skyrmion in AFM increases for either direction of the field perpendicular to the plane.
The effect is larger for the skyrmion in FM and it becomes unstable in a field of $B$=0.07~$|J|/\mu$. 
However, the increase in the size of the skyrmion in AFM leads to increased stability.

\begin{figure}[h!]
\centering
\includegraphics[width=\columnwidth]{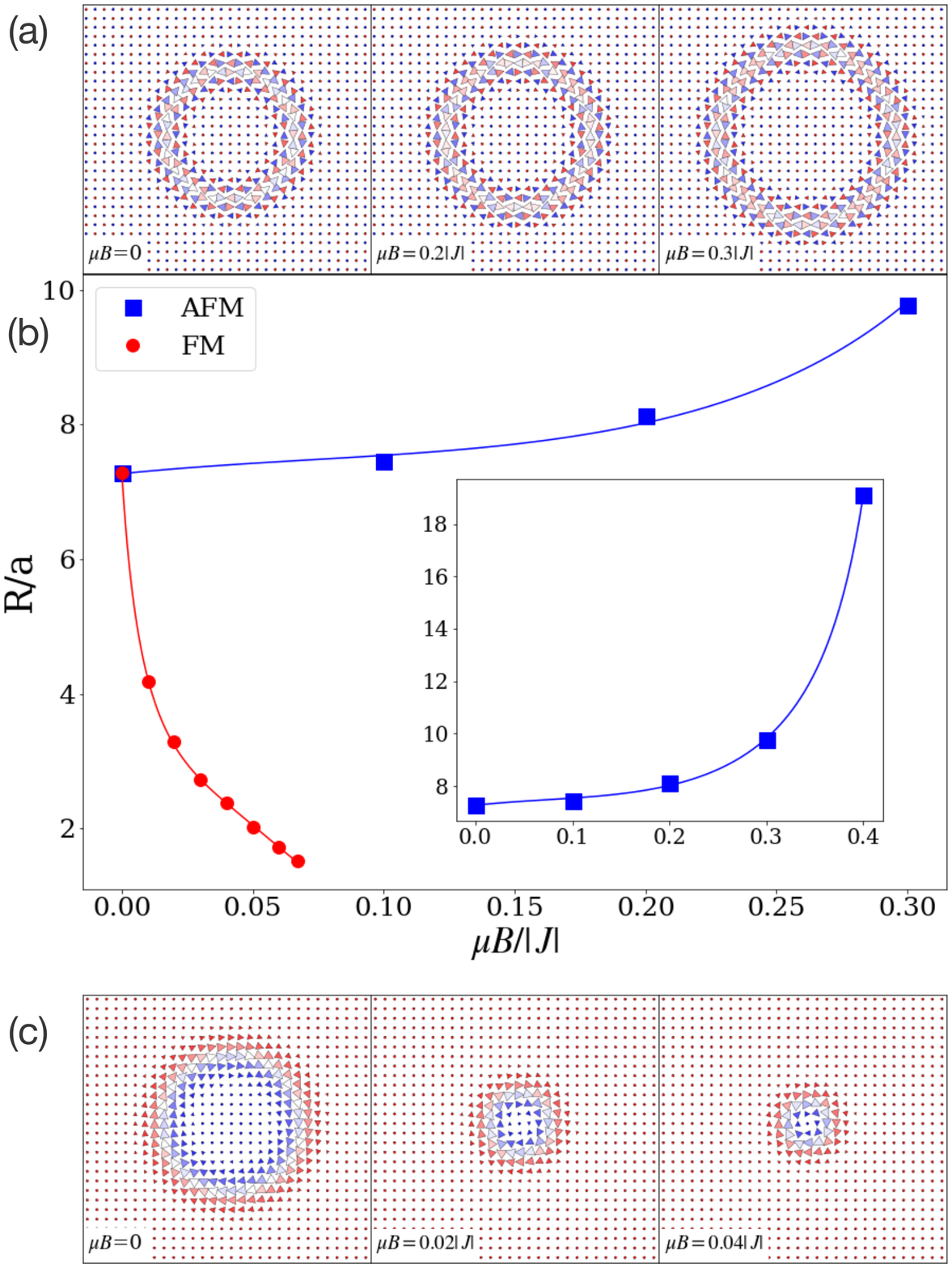}
\caption{
(a): 
 AFM skyrmion in a magnetic field of $B\, =\, 0, \  0.2 $ and  $0.3~ |J|/\mu$. 
 The color (red vs. blue) indicates the direction of the out-of-plane component of the 
 magnetic vector and the color intensity the magnitude.
(b): 
 Radius 
 (in units of lattice parameter, $a$) 
 of a skyrmion as a function of applied magnetic field in the AFM (blue) and the corresponding FM (red) where
 the sign of the exchange coupling parameter, $J$, and the Dzyaloshinskii-Moriya parameter, $D$, has been reversed. 
 In a field of 0.07 $|J|/\mu$, the FM skyrmion becomes unstable, but the AFM skyrmion is stable, even for significantly larger field 
 (see inset for extended scale).  
 While the radius of the skyrmion in an FM shrinks with applied field, it increases in an AFM.
(c): 
 FM skyrmion in a magnetic field of $B\, =\, 0, \  0.02$ and $0.04~|J|/\mu$. 
 Same color code as in top panel but smaller field.
}
\label{fig1}
\end{figure}

%

\subsection{Rates of annihilation processes} 

Fig. 2 shows MEPs for the collapse of a skyrmion in AFM and the corresponding FM with and without applied magnetic field.  
Not only are the relaxed skyrmions equivalent in the two materials in the absence of an applied magnetic field,
but this applies as well to the whole MEP. The activation energy for collapse is exactly the same. 
For parameters chosen here ($|J| \approx$ 31~meV, $K \approx $ 5~meV, $D \approx$ 11~meV, $\mu =$ 8~$\mu_B$) 
the activation energy for collapse of the skyrmion is 124 meV when $B$=0. 
This value of the magnetic moment in the 2D representation of the film could correspond to a 4 atomic layer film 
where each atom has a magnetic moment of 2~$\mu_B$.

\begin{figure}[h!]
\includegraphics[width=\columnwidth]{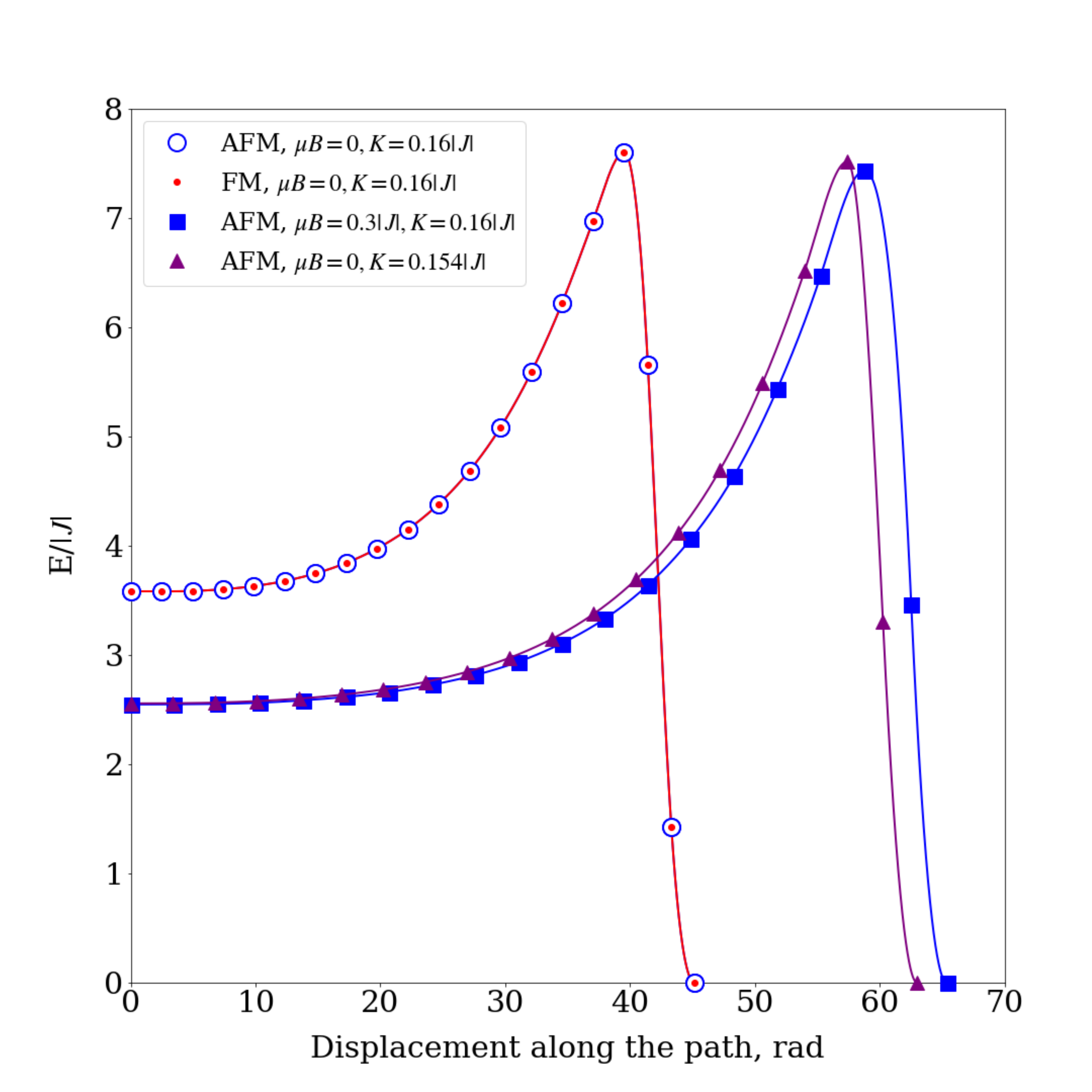}
\caption{
Minimum energy path for the collapse of a skyrmion in the AFM (blue) and in the corresponding FM (red) 
where the sign of the exchange coupling parameter, $J$, 
and the Dzyaloshinskii-Moriya parameter, $D$, have been reversed.
In the absence of an external magnetic field (dots and circles), the paths are identical for the two materials.
In a field of $B$=0.3 $|J|/\mu$, the activation energy for collapse is increased for a skyrmion in AFM  and the path lengthens (squares), 
but for the FM skyrmion the activation energy drops significantly as can be seen from the large increase in the rate shown in Fig.~3(a).
An application of a field of $B$=0.3 $|J|/\mu$ in the AFM has a similar effect as a
slight reduction of the anisotropy constant from $K$=0.16 $|J|$ to 0.154 $|J|$ in the absence of an applied field (triangles),
where again the path is the same for the two types of materials.
}
\label{fig2}
\end{figure}

When a magnetic field is applied, the skyrmion in AFM becomes more stable and the MEP for collapse longer.
Nearly the same effect on the path is obtained by modifying the anisotropy constant and setting the magnetic field to zero.
For example, the application of a field of $B$=0.3~$|J|/\mu$ is nearly equivalent to reducing the anisotropy constant from $K$=0.16 $|J|$ to 0.154 $|J|$ 
in accordance with Eqn.\,(2). The total displacement along the path and activation energy for annihilation are quite similar in the two cases. 
The field increases the radius of the skyrmion in AFM by a third and the activation energy for annihilation increases by about a third making the 
skyrmion in AFM more stable.




The value of the pre-exponential factor of the rate constant for skyrmion collapse is $1.0 \cdot 10^4 ~ s^{-1}$ for AFM
and $\nu_0 = 3.3 \cdot 10^3 ~s^{-1}$ for the corresponding FM. 

\begin{figure}[htbp]
\includegraphics[width=1.1\columnwidth]{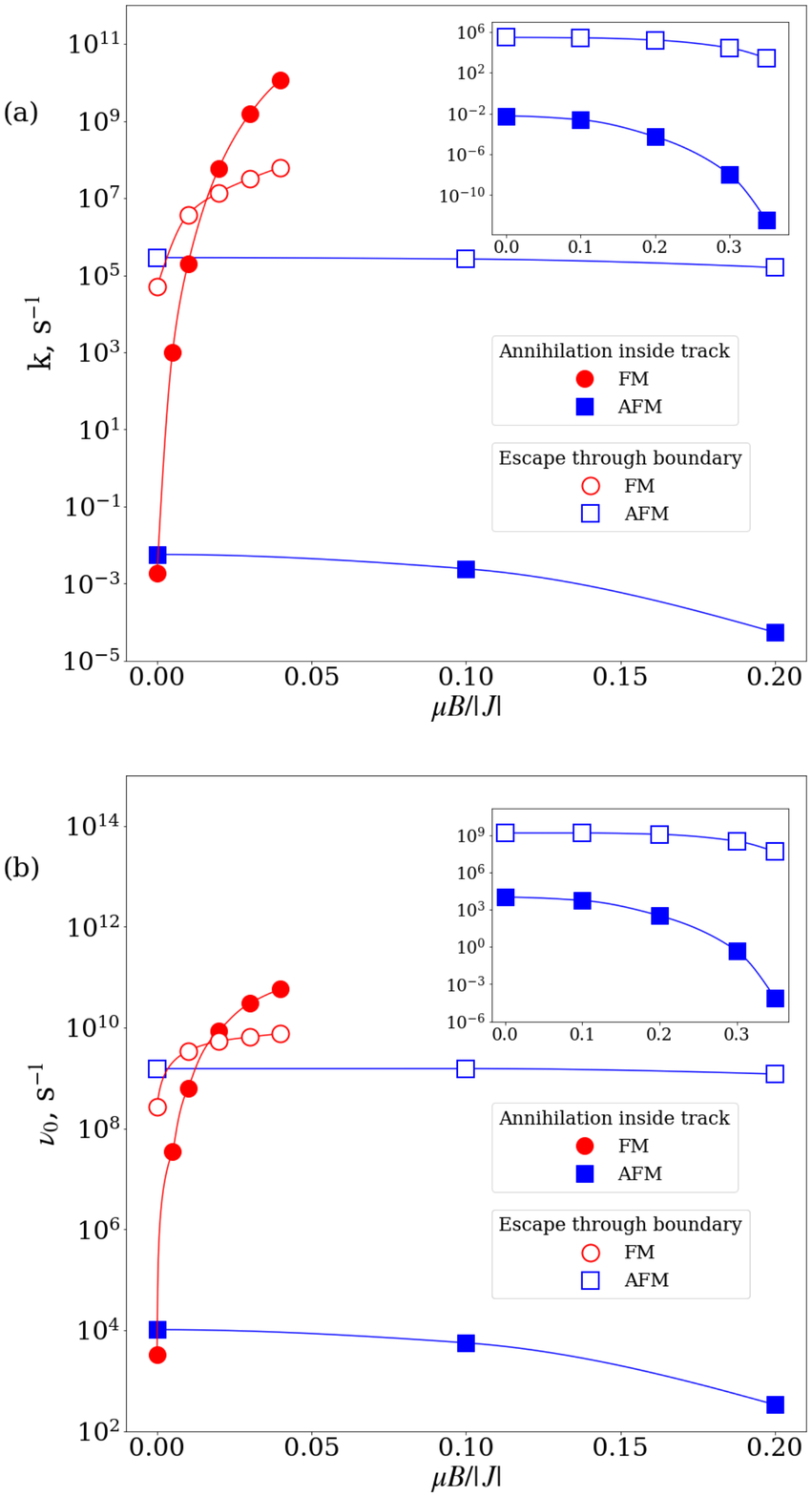}
\caption {
(a): 
Rate constant for the escape of a skyrmion through a boundary of the track and collapse inside the track 
as a function of applied magnetic field at a temperature of 
100 K.  
For a skyrmion in the AFM (blue) the effect of the field is weak, but for  
a skyrmion in the corresponding FM (red) the rate of annihilation increases dramatically as the strength of the applied field increases 
and a crossover occurs between collapse and escape, the former becoming more likely for large field. 
(b):  
Pre-exponential factor, $\nu_0$ for the rate constants illustrated in (a). 
A similar variation with the field strength is seen, a decrease for the AFM
but a large increase for the corresponding FM. 
Insets: Scale extended to larger field strength. 
}
\label{fig3}
\end{figure}

The effect of an applied field on the rate of collapse inside the track and escape through the boundary of the track
can be seen in Fig.~3. 
For the FM, the rate of both mechanisms increases strongly with the field. 
While escape has higher rate constant for small field, 
a crossover between the two mechanisms occurs at a field of $B$=0.02~$|J|/\mu$ and the rate constant for collapse inside the track 
becomes larger than escape through the edge for higher values of the field. 
The field has the opposite effect on the rate constants in the AFM as they become smaller when the field is increased, 
consistent with the increased activation energy shown in Fig.~2.
However, the cause of the drop in the value of the rate constants is not only because of the increase in the activation energy 
but also a decrease in the pre-exponential factor, $\nu_0$, as can be seen from Fig.~3(b).  
This reflects a decrease in the entropy of the transition state with respect to the entropy of the initial state. 
It can be understood from the fact that the larger the skyrmion is, the larger its vibrational entropy is.
As the field increases, the skyrmion in AFM becomes larger, thus increasing the entropy of the initial state. 
But, at the transition state the skyrmion is reduced to a critical size that depends less on the applied field. 
So, the entropy of the transition state increases less with the field than the entropy of the initial state.

A more likely annihilation mechanism at zero magnetic field for this set of parameter values is an escape of the skyrmion through the edge of the track. 
Even if the relative number of sites adjacent to the edge and the number of sites in the interior of the track is taken into account, 
the probability of escape through the edge of the track is much more likely than collapse. This means that additional effort, such as insertion of material with other 
magnetic characteristics at the boundary, or pinning of skyrmions by defects such as a narrow strip with a different materials parameter\cite{Stosic2017} 
are necessary to keep the skyrmion inside the track. 

While the activation energy for collapse is exactly the same for a skyrmion in the AFM and in the corresponding FM when there is no applied field,
the pre-exponential factor in the rate expression is different. The reason is that the Landau-Lifschitz dynamics of the skyrmion in the transition state is different
in the two materials.
Fig. 4 shows the contributions of the various eigenmodes ${\bf e}_k$: $f_k={\bf e}_k\cdot({\bf S}^{sp}\times {\bf e})$ 
to the factor $\lambda$ in the pre-exponential (see Eqn.~(5)).  
The eigenvectors ${\bf e}_k$ are numbered in increasing order of  the corresponding eigenvalues. 
The calculated results show that small eigenvalues give the major contribution for
the transition state for skyrmion collapse in the FM (note the logarithmic scale), whereas the 
full range of eigenvalues contributes in the AFM. 
As a result, the value of the pre-exponential factor for the collapse of a skyrmion in the AFM and the corresponding FM turn out to differ. 

\begin{figure}[h!]
\includegraphics[width=\columnwidth]{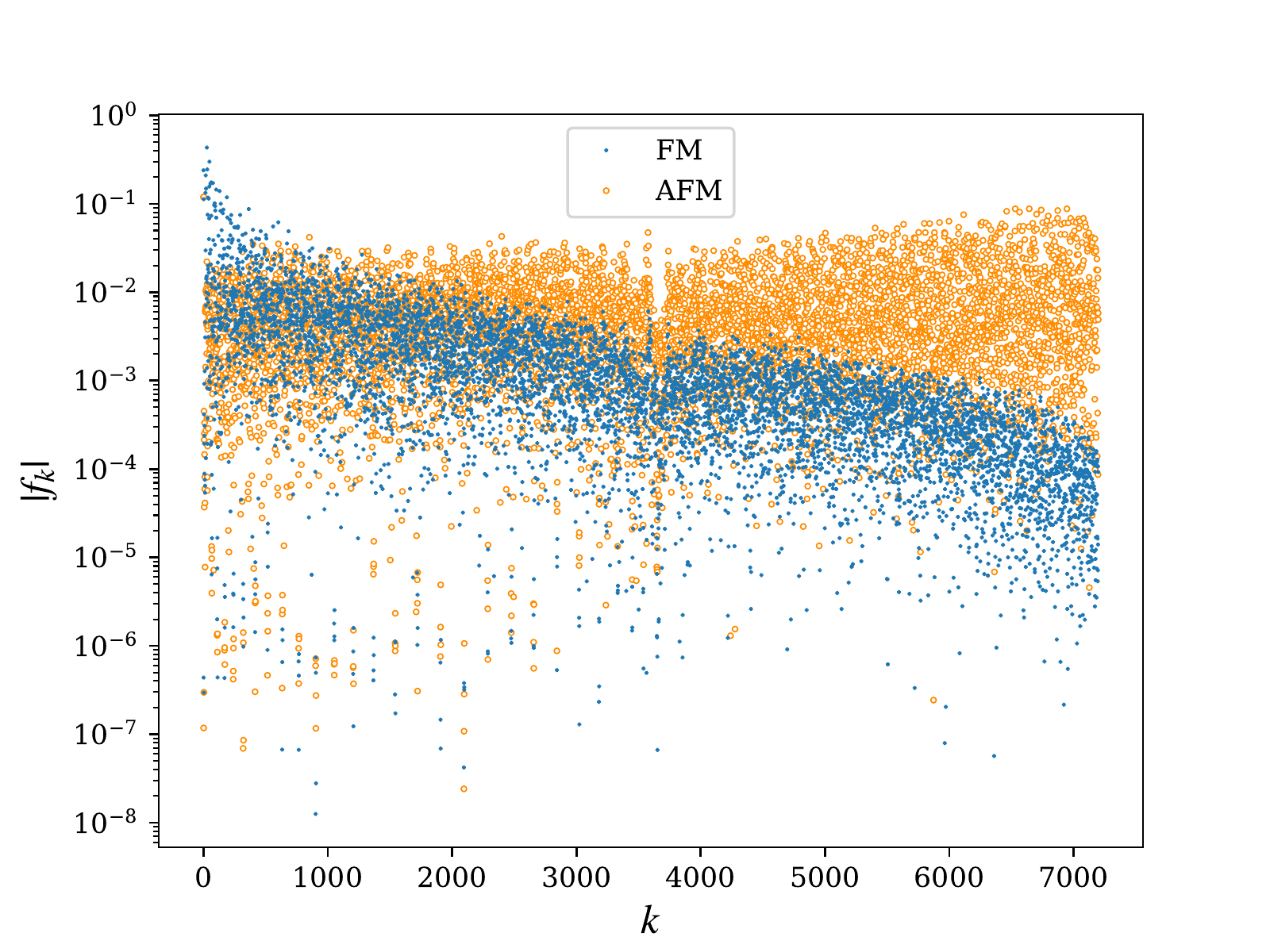}
\caption{
The contribution $f_k$ of the $7199$ positive eigenmodes of the Hessian at the saddle point for skyrmion collapse to the factor $\lambda$ in 
the pre-exponential factor of the rate constant (see Eqn.~(5)).
The x-axis gives the number of the eigenvalue arranged in increasing order. 
Small eigenvalues give the major contribution for the transition state of skyrmion collapse in the FM (note the logarithmic scale), whereas the 
full range of eigenvalues contributes in the AFM. 
The dynamical contribution to the pre-exponential factor is different for the two because the Landau-Lifshitz dynamics are different.
}
\label{fig4}
\end{figure}


\begin{figure}[h!]
\includegraphics[width=\columnwidth]{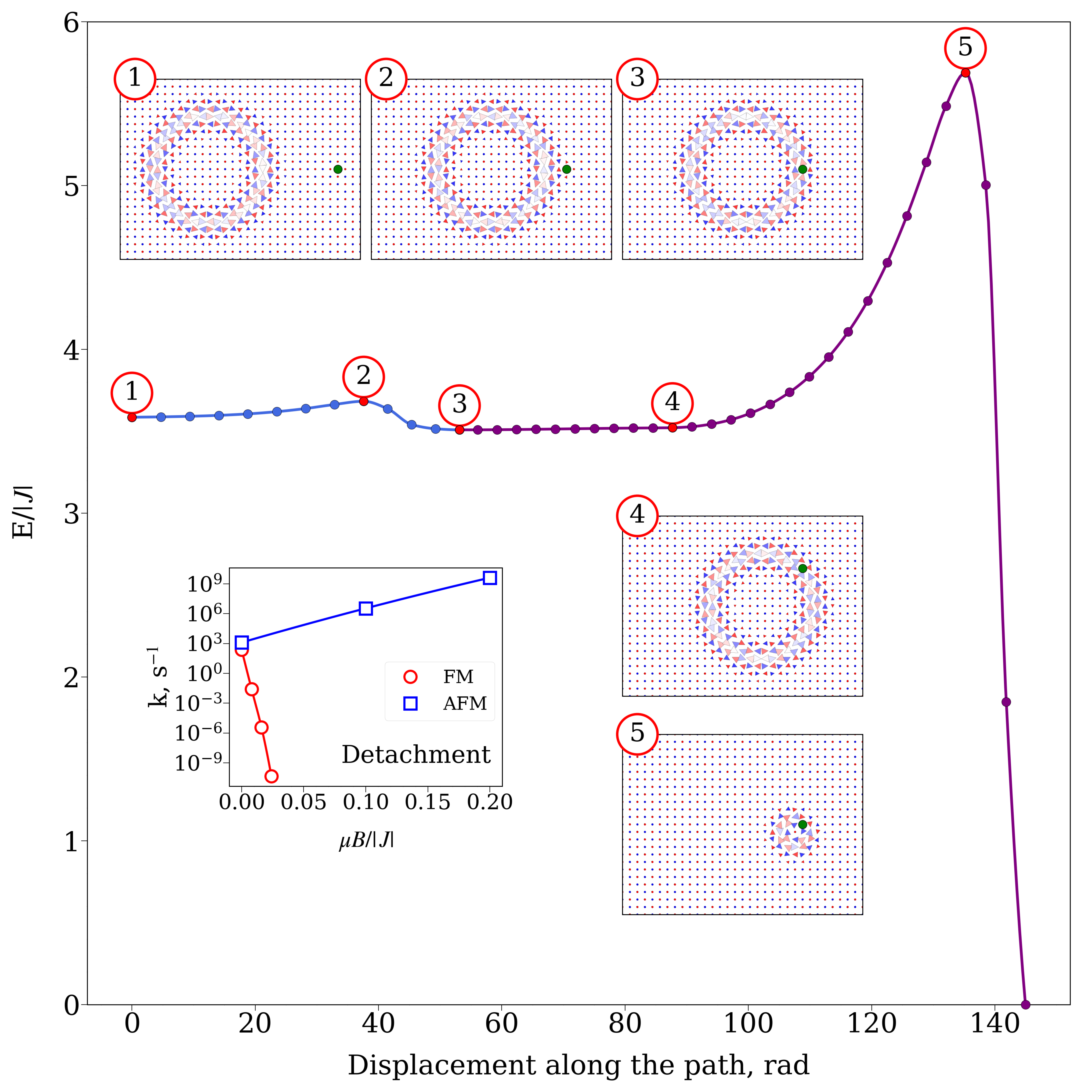}
\caption{
Minimum energy path for the attachment of a skyrmion to a non-magnetic impurity atom in the AFM and subsequent collapse there.
The lowest energy spin configuration is when the defect is located in a region of the skyrmion where the magnetic moments have no out-of-plane component. 
The skyrmion can rotate without changing the energy as long as the impurity remains at the same distance from the center (compare points 3 and 4). 
Inset graph:
Rate constant for the detachment from the impurity as a function of applied magnetic field strength at $T$=3 K. 
The skyrmion detaches more readily from an impurity in the AFM than in the corresponding FM, 
and this difference becomes greater as the field strength increases.
In the AFM, the rate constant for dissociation from the impurity is always larger than the rate constant for collapse even for large applied field.
Insets 1-5: Configurations of the magnetic vectors at the various labeled points along the minimum energy path.
}
\label{fig5}
\end{figure}

\subsection{Non-magnetic impurity}

An important issue for practical applications of magnetic skyrmions is the effect of impurities in the magnetic material.
The effect of a non-magnetic impurity on the stability and pinning of a skyrmion in FM has been presented previously.
\cite{Uzdin18-ph_b}
Fig.~5 shows the minimum energy path for the attachment of a skyrmion in AFM to a non-magnetic substitutional 
atom and subsequent collapse of the skyrmion. 
There is a small activation energy barrier for attaching to the defect, but once that has been overcome the energy drops 
below that of a free skyrmion. The optimal configuration is obtained when the non-magnetic atom is located in the 
region where the magnetic moments have no out-of-plane component, see Fig.~5.
This defines the optimal distance between the impurity and the center of the skyrmion. The skyrmion
can rotate freely around the impurity as long as the distance to the impurity remains constant. 
This motion involves no change in the energy of the system (see the long stretch between points 3 and 4 in Fig.~5). 

There is still a large probability that the skyrmion detaches from the impurity because the energy barrier is rather small. 
The impurity, however, also increases greatly the probability of collapse as the activation energy drops to a half compared
to that of a free skyrmion, as can be seen by comparing Fig.~5 with Fig.~2. 
The impurity stabilizes the contracted skyrmion at the transition state more than the relaxed skyrmion in the initial state, 
thereby lowering the activation energy.
Correspondingly, the lifetime drops if the skyrmion is bound to the impurity. 
An applied field has a beneficial effect for a skyrmion in the AFM by increasing the rate of detachment (see inset in Fig.~5) as well as
by increasing the energy barrier for collapse, as shown in Fig.~2.


\section{Discussion}

Skyrmions in AFMs and corresponding FMs can be described by similar energy surfaces. 
Without an applied magnetic field the energy surfaces coincide exactly. 
The inclusion of an applied magnetic field for the AFM material is equivalent to a change in in the anisotropy in the corresponding FM material without
applied field. 
Therefore, the activation energy for annihilation and the entropy contribution to the pre-exponential factor $\Omega_0$ in the transition rate is exactly the same 
for  AFM and corresponding FM skyrmions. 
However, the rates are different because the dynamics through the transition state, given by the Landau-Lifschitz equation, 
is different. The calculated results presented here illustrate this for an example system.
This result is in agreement with recently reported mean switching time by coherent rotation of AFM nanoparticles in the superparamagnetic limit, 
which have shown different pre-exponential factor for AFM and FM single-domain particles.\cite{Rozsa19}

The effect of a non-magnetic impurity is found to be quite different for a skyrmion in the AFM than in the corresponding FM, 
especially in the presence of an applied field.
Recently, dynamical modeling of the movement of the skyrmion inside a track and its collision with hole defects was reported.\cite{Silva19} 
Depending on the strength of the applied current and the type of collision, a skyrmion in an AFM can be captured, scattered or completely destroyed by the hole. 
Evidently, these phenomena are connected with the energy barriers and pre-exponential factors presented above.

The key question is whether it is possible to find an AFM where skyrmions are stable enough at room temperature. 
The equivalence of the energy surfaces for skyrmions in AFM and in the corresponding FM gives hope that this is possible, 
because stable skyrmions in FM materials at room temperature have been reported experimentally.\cite{soumyanarayanan2017,Fert17,Everschor-Sitte18}
FM skyrmions that have been studied experimentally, even those which exist in ultra thin magnetic films on a surface of a heavy metal,\cite{Heinze2011} 
contain several magnetic atomic layers. If the thickness of the film is much less than the size of the skyrmion then the structure can be considered as a quasi 
2D system. Such skyrmions are tubes with the same magnetic structure in each atomic layer.\cite{Sohn19,Vlasov20} 
In this approach, a column of magnetic moments is  
represented by a macrospin with
effective parameters within the 2D model.
Parameters of the Hamiltonian in Eqn.~(1), such as $J, ~K, ~D$ and $ \mu$, are then to a first approximation proportional to the thickness of the magnetic film whereas the skyrmion radius and in-plane structure do not depend on the thickness.  
Evidently, if the sign of the exchange and DM interaction between magnetic moments in neighboring layers is changed as well as the directions of the magnetic moments, the same energy surface for AFM and FM is obtained as in the 2D case. 
For AFM, the vector $L$ should be considered instead of local moments, but the effective parameters of the Hamiltonian will, 
to first approximation, 
be proportional to the AFM film thickness as in the case of FM films.

Density functional theory (DFT) calculations have given estimates of the parameter $J$ characterizing the exchange interaction between nearest 
neighbor magnetic moments around 10 meV.\cite{Malottki2017} 
If the magnetic film contains several atomic layers, this parameter can be proportionally larger. 
As for FM materials, the parameter values for an AFM film scale with the thickness of the magnetic layer. 
The value of the exchange parameter $J$, anisotropy 
parameter $K$,  and even magnetic moment $\mu$ (a macrospin representation) can be proportional to the number of magnetic layers. 
Note that the effective magnetic moment in the AFM case corresponds to the value of staggered magnetization {\bf L}. 
Thus the effective values of the parameters in the 2D Hamiltonian Eqn.~(1) can be much larger than 
values of the interactions between local magnetic moments calculated by the DFT method.\cite{Bessarab2018,Malottki2017} 
Note, however, that the DM interaction in a quasi-2D system is associated with the presence of a heavy metal 
with large spin orbit interaction and this effect decays rapidly away from the interface. 
The DM parameter is then not proportional to the thickness of the film.
Nevertheless recent calculations\cite{Ma19} show that in an amorphous ferrimagnetic GdCo film on Pt, 
the skyrmion assumes a columnar configuration that extends uniformly across the film thickness of 10 nm 
despite having near zero DM interaction far away from the interface. 
Effective DM interaction is still enough to stabilize the skyrmion tube.

The properties of a skyrmion in an AFM is related here to the properties of a skyrmion in a corresponding FM by a transformation of the relevant 
parameters in the Hamiltonian and the orientation of the spins. Similar analysis can be carried out for other topological magnetic textures.
For example, the properties of an antiskyrmion in a FM have been analyzed by spin transformation that converts it to a skyrmion in the FM and corresponding
changes in the parameters in the Hamiltonian if there are no spin-orbit and spin-transfer torques. This means that the lifetime of the antiskyrmion is exactly the same as for the skyrmion if the DM vector is transformed accordingly.  Moreover, by applying this transformation to the Landau-Lifshitz-Gilbert equation 
of motion one can predict the direction of the current for which the angle between current and antiskyrmion movement equals zero.\cite{Potkina19}

The data that support the findings of this study are available from the 
corresponding 
author upon a reasonable request.


\begin{acknowledgments}
This work was supported by the Icelandic Research Fund, the Research Fund of the University of Iceland, the Russian Foundation of Basic Research (grants RFBR 18-02-00267 and 19-32-90048) and the Foundation for the Advancement of Theoretical Physics and Mathematics ``BASIS'' under Grant No. 19-1-1-12-1. 
\end{acknowledgments}


\end{document}